\documentclass{JHEP3}
\usepackage{amsmath}
\input epsf
\usepackage{epsfig}
\usepackage{amssymb}
\usepackage[active]{srcltx}

\newcommand{\bea}{\begin{eqnarray}}
\newcommand{\eea}{\end{eqnarray}}
\newcommand{\bean}{\begin{eqnarray*}}
\newcommand{\eean}{\end{eqnarray*}}
\newcommand{\nn}{\nonumber \\}

\def\W #1{\widetilde{#1}}

\def\braket#1{\left\langle #1 \right\rangle}

\def\gb #1{ \left\langle #1 \right]}

\def\a{{\alpha}}

\def\b{{\beta}}

\def\la{\lambda}
\def\eps{\epsilon}

\def\vev{\braket}

\def\bvev#1{\left[ #1 \right]}
\def\Spaa{\vev}
\def\Spbb{\bvev}
\def\Spab{\gb}

\def\Label#1{\label{#1}%
  \smash{\hbox to0pt{\raise1ex\hbox{\tiny[#1]}\hss}}}

\title{$U(1)$-decoupling, KK and BCJ relations in $\mathcal{N}=4$ SYM}
\author{ Yin Jia$^{\dagger}$, Rijun Huang$^{\dagger}$,  Chang-Yong Liu$^\diamond$\footnote{email address: lcy@itp.ac.cn}~~~~\\
$^{\dagger}$Zhejiang Institute of Modern Physics, Physics Department, Zhejiang University, Hangzhou, China\\
$^\diamond$Center of Mathematical Science, Zhejiang University,
Hangzhou, China\\
}
\date{\today}
\abstract{By using the BCFW recursion relation of $\mathcal{N}=4$
super Yang-Mills theory, we proved the color reflection,
$U(1)$-decoupling, Kleiss-Kuijf and Bern-Carrasco-Johansson
relations for color-ordered amplitudes of $\mathcal{N}=4$ SYM
theory. This proof verified the conjectured BCJ relations of matter
fields. The proof depended only on general properties of
super-amplitudes. We showed also that color reflection relation and
$U(1)$-decoupling relation are special cases of KK relations.}

\begin{document}

%%%%%%%%%%%%
\section{Introduction}
%%%%%%%%%%%%%%%%%%

The calculation of scattering amplitudes is important in quantum
field theory. Generally one can use Feynman diagrams to compute the
scattering amplitudes, but the number of Feynman diagrams increases
dramatically with the increasing of external particles, therefore it
is hard to handle the calculation both theoretically and
practically. Many efficient methods have been proposed to solve this
problem, especially under the practical demanding of newly particle
experiments, such as LHC in Geneva. Among these methods there is
BCFW on-shell recursion relation\cite{Britto:2004ap, Britto:2005fq},
which was inspired by Witten's twistor program \cite{Witten:2003nn}.
BCFW recursion relation is a powerful method to obtain simpler and
more compact expressions for tree-level amplitudes. Besides this, it
is also a very powerful tool to prove many useful relations of
amplitudes. The supersymmetric BCFW recursion relation has also been
written down  recently \cite{Bianchi:2008pu, Brandhuber:2008pf,
ArkaniHamed:2008gz, Elvang:2008na} and has been applied to many
places such as \cite{Drummond:2008cr}.

Through the calculation of amplitudes, a number of useful relations
have been found for color-order tree amplitudes of gluons. These
include the color-order reflection relation, the $U(1)$-decoupling
relation, the Kleiss-Kuijf relations \cite{Kleiss:1988ne} and the
BCJ-relations \cite{Bern:2008qj}. The Kleiss-Kuijif relations reduce
the number of independent color-ordered amplitudes of $n$-point
gluons to $(n-2)!$. The BCJ-relations reduce the number of
independent color-ordered amplitudes further to $(n-3)!$. The KK
relations have been proved by field theory in \cite{DelDuca:1999rs}
and again proved beautifully, together with BCJ-relations, using
string theory method in
\cite{BjerrumBohr:2009rd,Stieberger:2009hq}(see further works
\cite{Tye:2010dd,BjerrumBohr:2010zs,Mafra:2009bz}). Another proof of
all these relations has been given in \cite{Feng:2010my} using BCFW
recursion relation, where only general properties of scattering
amplitudes in S-matrix program are used.

In this short note we extend the work of \cite{Feng:2010my} to
$\mathcal{N}=4$ SYM theory. More accurately, starting from the color
reflection relation of three-point super-amplitudes, which can be
directly verified in S-matrix program
(\cite{Benincasa:2007xk,S-matrix}), plus supersymmetric BCFW
recursion relation, we will prove following four properties: (1)
color reflection relation for general $n$; (2) $U(1)$-decoupling
relation; (3) KK relations and (4) BCJ relations. Since we can
expand the super-field into on-shell component fields through the
$\eta^A$ series, after getting the relations of super-amplitudes of
$\mathcal{N}=4$ SYM theory we can recover the relations of gluon
amplitudes and BCJ relations of matter fields from super-amplitude
expanding. This verifies the conjecture in
\cite{Sondergaard:2009za}.

The paper is organized as follows, in section two we briefly review
some basic facts of $\mathcal{N}=4$ SYM theory, including the BCFW
recursion relation of SYM version. In section three and four we
prove the color reflection relation and $U(1)$-decoupling relation.
In section five, we prove the KK relations. Before giving the
general proof we show that color reflection relation and
$U(1)$-decoupling relation are just special cases of KK relations.
In section six, we show that there is a primary relation in general
formula of BCJ relations and other BCJ relations can be generated
from this primary one combined with KK relations. Conclusion and
discussions are presented in the last section.

%%%%%%%%%%%%%%%%%
\section{Review}
%%%%%%%%%%%%%%%%%
It is well known that tree-level amplitude of pure gluons is
identical to the one obtained from ${\cal N}=4$ supersymmetric
Yang-Mills theory. For ${\cal N}=4$ SYM theory, we can group all
components into following on-shell superfield
\cite{ArkaniHamed:2008gz,Drummond}
\bea \Phi(p,\eta) & = & G^+(p)+\eta^A \psi_A^+(p) +{1\over 2}
\eta^A\eta^B S_{AB}(p)+{1\over 3!}\eta^A\eta^B \eta^C \eps_{ABCD}
\psi^{D-}(p)+{1\over 4!}\eta^A\eta^B \eta^C \eta^D\eps_{ABCD}
G^-(p)\nonumber \\ \eea
with Grassmann coordinates $\eta^A, A=1,2,3,4$. Using superfield,
the amplitude can be written as functions of $(\la_i,\W\la_i,
\eta_i^A)$. For example, the $n$-point super-MHV amplitude is given
by Nair's formula\cite{Nair:1988bq} as
\bea {\cal A}(\la,\W\la, \eta) = {\delta^4(\sum_i \la_i
\W\la_i)\delta^8(\sum_i \la_i \eta_i^A)\over
\Spaa{1~2}\Spaa{2~3}...\Spaa{n~1}}~,~~~\label{SUSY-MHV}\eea
where the Grassmann variables $\eta_i$ appear in super-delta
function to indicate the super energy-momentum conservation
$\sum_i\lambda_i\eta_i=0$. To obtain the corresponding scattering
amplitudes for various components we just need to expand above
expression with respect to $\eta^A$. More concretely, after
expanding into series of $\eta^A$ we get
\bea {\cal A} = \sum_{a_1.. a_n=0}^4 {\cal A}_{a_1... a_n}
\prod_{i=1}^n\eta_i^{a_i}~,~~~\label{SUSY-exp}\eea
where ${\cal A}_{a_1... a_n}$ gives amplitude with particular field
configuration indicated by $(a_1,a_2,...,a_n)$. In this case, if we
have the relations of super-amplitudes for $\mathcal{N}=4$ SYM
theory, we can expand super-amplitudes into component fields by
$\eta^A$ series, then we obtain relations for gluon and matter
fields.

We will prove many relations of super-amplitudes by induction
starting from relations of three-point super-amplitudes. By direct
verification we know that the general three-point super-amplitudes,
including the MHV and $\overline{\mbox{MHV}}$ super-amplitudes, have
the property $\mathcal{A}(1,2,3)=-\mathcal{A}(3,2,1)$. We can
observe this property directly from the component amplitudes. For
components of pure gluon we have
\bea A(1^-,2^-,3^+)={\Spaa{1~2}^3\over \Spaa{2~3}\Spaa{3~1}}~, \quad
A(1^+,2^+,3^-)={\Spbb{1~2}^3\over \Spbb{2~3}\Spbb{3~1}}~.~\eea
Let us take $A(1^-,2^-,3^+)$ for example, after using cyclic
relation we have
\bea A(3^+,2^-,1^-)=A(2^-,1^-,3^+)={\Spaa{2~1}^3\over
\Spaa{1~3}\Spaa{3~2}}=-{\Spaa{1~2}^3\over
\Spaa{2~3}\Spaa{3~1}}=-A(1^-,2^-,3^+)~,\eea
where we have used the property that $\Spaa{i~j}=-\Spaa{j~i}$. The
same argument applies to $\overline{\mbox{MHV}}_3$ where
$\Spbb{i~j}=-\Spbb{j~i}$ is used, thus we have $A(1,2,3)=-A(3,2,1)$.
For components of two-fermion and one gluon we have
\bea A(1_f^-,2_g^-,3_f^+)={\Spaa{1~2}^3\Spaa{3~2}\over
\Spaa{1~2}\Spaa{2~3}\Spaa{3~1}}~, \quad
A(1_f^-,2_g^+,3_f^+)={\Spbb{1~2}^3\Spbb{3~2}\over
\Spbb{1~2}\Spbb{2~3}\Spbb{3~1}}~.\eea
Similarly we can write down the amplitudes of $A(3_f^+,2_g^-,1_f^-)$
and $A(3_f^+,2_g^+,1_f^-)$, by comparing to the ones given above,
and after paying attention to the case that there is one minus sign
from $\eta_i$ exchanging, we can see again the above relations
preserved. The left case is amplitudes of two-scalar and gluon,
\bea A(1_s,2_g^-,3_s)={\Spaa{1~2}^2\Spaa{3~2}^2\over
\Spaa{1~2}\Spaa{2~3}\Spaa{3~1}}~, \quad
A(1_s,2_g^+,3_s)={\Spbb{1~2}^2\Spbb{3~2}^2\over
\Spbb{1~2}\Spbb{2~3}\Spbb{3~1}}~. \eea
This is similar to the pure gluon case. We reverse the color order
$(1,2,3)\to(3,2,1)$ and would see that $A(1,2,3)=-A(3,2,1)$. Thus we
confirm the relation of three-point super-amplitudes
\bea \mathcal{A}(1,2,3)=-\mathcal{A}(3,2,1)~.~~~~\label{3pbasic}\eea
Relation (\ref{3pbasic}) can be considered as color reflection
relation or $U(1)$-decoupling relation of three-point. It is the
fundamental relation and starting from this relation we can prove
relations of general amplitudes by induction.

In order to prove the $U(1)$-decoupling, KK and BCJ relations in
$\mathcal{N}=4$ SYM theory we need also the BCFW recursion relation
of supersymmetric version\cite{Bianchi:2008pu, Brandhuber:2008pf,
ArkaniHamed:2008gz, Elvang:2008na}, which is given by
\begin{eqnarray}
&&\mathcal{A}(\{\eta_1,\lambda_1,\tilde{\lambda}_1\},\{\eta_2,\lambda_2,\tilde{\lambda}_2\},\eta_i)=\nonumber\\
&&\sum_{L,R}\int
d^{4}\eta~\mathcal{A}(\{\eta_1(z_P),\lambda_1(z_P),\tilde{\lambda}_1\},\eta,\eta_L){1\over
P^2}\mathcal{A}(\{\eta_2,\lambda_2,\tilde{\lambda}_2(z_P)\},\eta,\eta_R)\label{n4bcfw}\end{eqnarray}
with (1,2)-shifting. The amplitudes in terms of BCFW expansion are
tree level on-shell super-amplitudes of lower point, while $\eta_L$
and $\eta_R$ are Grassmann variables attached to left and right
sub-super-amplitudes. Note that $\eta_1$ has also been shifted,
$\eta_1(z)=\eta_1+z\eta_2$, while others are unchanged, to keep the
super energy-momentum conservation $\sum_i \lambda_i(z)\eta_i(z)=0$
if we shift $\lambda_1=\lambda_1+z\lambda_2$. The super-space
integration is over shifted $\eta=\eta(z)$, and in $\mathcal{N}=4$
SYM theory, this integration can always be carried out by
super-delta function from super-amplitudes, without doing actual
calculation.

One important observation of (\ref{n4bcfw}) is that $\hat{P}(z)$
depends only on the sum of momenta and has nothing to do with color
order. The same fact exists for $\hat{\eta}(z)$, which is decided by
the manner of $\eta_1$ shifting and also has nothing to do with the
color order. In this case we can group terms of the same channel in
BCFW expansion, which usually have different color-order
sub-super-amplitudes, without changing the $\eta(z)$ integration.
Another important property is that when taking the $(i,j)$-shifting,
super-amplitudes of $\mathcal{N}=4$ SYM have a good behavior ${\cal
A}^{{\cal N}=4}(z)\to 0$ with $z\to \infty$ no matter which
helicities of the shifted momenta are\cite{ArkaniHamed:2008gz},
i.e., there is no boundary contribution. The boundary behavior is
very important when applying the BCFW recursion relation, and there
are many works on the boundary behavior\cite{ArkaniHamed:2008yf,
Vaman:2005dt,Draggiotis:2005wq,Benincasa:2007qj,Cheung:2008dn,Feng:2009ei,Feng:2010ku}.
Luckily in $\mathcal{N}=4$ SYM theory we do not have bad deformation
at all, thus we do not need to consider the details of
$(i,j)$-shifting. We should also emphasize that for $\mathcal{N}=4$
super-amplitudes, when the shifted momenta are adjacent, we have
${\cal A}^{\mathcal{N}=4}(z)\to 1/z$ when $z\to \infty$, while the
shifted momenta are not adjacent, we have ${\cal
A}^{\mathcal{N}=4}(z)\to 1/z^2$ when $z\to \infty$. This fact helps
a lot when proving the BCJ relations in gauge theory, as well as in
gravity where they are just the bonus relations
\cite{ArkaniHamed:2008gz,Spradlin:2008bu}.

%%%%%%%%%%%%%%%%
\section{The color reflection relation}
%%%%%%%%%%%%%%%%

The general expression for color reflection relation of
super-amplitudes can be written as
\bea
\mathcal{A}(1,2,\ldots,n)=(-)^n\mathcal{A}(n,n-1,\ldots,1)~.~~~~\label{creflection}\eea
Since color reflection relation of three-point super-amplitudes is
satisfied, we want to prove the general case by induction starting
from three-point case. Using the BCFW recursion relation of
$\mathcal{N}=4$ SYM theory and (1,n)-shifting we have
\bean \mathcal{A}(1,2,\ldots,n) &=&\int d^4\eta_{\hat{P}}
\sum_{i=2}^{n-2}
\mathcal{A}(\hat{1},\ldots,i,-\hat{P}_i;\eta_{\hat{P}}){1\over
P_i^2}\mathcal{A}(\hat{P}_i,i+1,\ldots,\hat{n};\eta_{\hat{P}})~,\eean
then we use $(i+1)$ and $(n-i+1)$-point color reflection relations
to re-write these sub-super-amplitudes as
\bean &&\mathcal{A}(\hat{1},\ldots,i,-\hat{P}_i;\eta_{\hat{P}})=
(-)^{i+1}\mathcal{A}(-\hat{P}_i,i,\ldots,\hat{1};\eta_{\hat{P}})~,\\
&&\mathcal{A}(\hat{P}_i,i+1,\ldots,\hat{n};\eta_{\hat{P}})=
(-)^{n-i+1}\mathcal{A}(\hat{n},\ldots,i+1,\hat{P}_i;\eta_{\hat{P}})~,
\eean
where variable $z$ of shifted momenta in sub-super-amplitudes has
been replaced by $z_P$ so that every momentum in
sub-super-amplitudes is on-shell, and we are able to use color
reflection relation of lower-point. Putting sub-super-amplitudes of
the left hand side on the right hand side and vice versa, we have
\bean \mathcal{A}(1,2,\ldots,n)&=&\int d^4\eta_{\hat{P}}
\sum_{i=2}^{n-2}(-)^{n-i+1}\mathcal{A}(\hat{n},\ldots,i+1,\hat{P}_i;\eta_{\hat{P}})
{1\over
P_i^2}(-)^{i+1}\mathcal{A}(-\hat{P}_i,i,\ldots,\hat{1};\eta_{\hat{P}})\\
&=&(-)^n\mathcal{A}(n,n-1,\dots,1)~.\eean
In this step we have used the factor that $\eta_{\hat{P}}$ depends
on the sum of $\eta_i$ and does not change under color reflection,
so that we can write these terms back to get color reversed
super-amplitudes and finish the proof.

%%%%%%%%%%%%%%%%%%%%%%%%%%
\section{The $U(1)$-decoupling relation}
%%%%%%%%%%%%%%%%%%%%%%%%%%

%%%%%%%%%%%%%%%%%%%%%
\subsection{The four-point case}
%%%%%%%%%%%%%%%%%%%%%%

Let us start with the simplest case of $U(1)$-decoupling relation,
which is given by
\bea \mathcal{A}(1,2,3,4)+
\mathcal{A}(1,3,4,2)+\mathcal{A}(1,4,2,3)=0~.~~~~\label{U1-4}\eea
Let us take $(1,2)$ to do the shifting. As mentioned before, in
$\mathcal{N}=4$ SYM theory, no matter what the helicity
configuration is, there is always a good deformation for given pair
$(i,j)$. When doing the BCFW expansion of SYM version we should also
shift corresponding $\eta_i$ to keep the super-energy-momentum
conservation. The $\eta$ shifting depends on the way taking
$\Spab{1|2}$-shifting or $\Spab{2|1}$-shifting, but since the
super-space integration is only formally kept in the steps of
demonstration, it is not necessary caring about the details of
shifting. With the choice of $(1,2)$ pair we have following
contributions for various amplitudes. Firstly for
$\mathcal{A}(1,2,3,4)$ we have
\bea \mathcal{A}(1,2,3,4) & = & \int
d^4\eta_{\hat{P}}\mathcal{A}(\hat{2},3,-\hat{P}_{23};\eta_{\hat{P}}){1\over
P^2_{23}}\mathcal{A}(\hat{P}_{23},4,\hat{1};\eta_{\hat{P}})~.~~~~\label{A1234}\eea
Then for $\mathcal{A}(1,3,4,2)$ we have
\bea \mathcal{A}(1,3,4,2) & = & \int
d^4\eta_{\hat{P}}\mathcal{A}(\hat{1},3,-\hat{P}_{13};\eta_{\hat{P}}){1\over
P^2_{13}}\mathcal{A}(\hat{P}_{13},4,\hat{2};\eta_{\hat{P}})~,~~~~\label{A1342}\eea
and finally for $\mathcal{A}(1,4,2,3)$ we have two pole
contributions given as
\bea \mathcal{A}(1,4,2,3) & = & \int
d^4\eta_{\hat{P}}\left[\mathcal{A}(\hat{1},4,\hat{P}_{23};\eta_{\hat{P}}){1\over
P^2_{23}}\mathcal{A}(-\hat{P}_{23},\hat{2},3;\eta_{\hat{P}})+\mathcal{A}(3,\hat{1},-\hat{P}_{13};\eta_{\hat{P}}){1\over
P^2_{13}}\mathcal{A}(\hat{P}_{13},4,\hat{2};\eta_{\hat{P}})\right]~.~~~~\label{A1423}\eea
Now let us compare various terms. The first term of (\ref{A1423}) is
almost same as the one in (\ref{A1234}) with only difference in the
order of the factor $\mathcal{A}(\hat{1},4,\hat{P}_{23})$ v.s.
$\mathcal{A}(4,\hat{1},\hat{P}_{23})$, and has the same
sub-super-amplitude $\mathcal{A}(-\hat{P}_{23},\hat{2},3)$, thus the
sum of these two terms is proportional to
$\mathcal{A}(-\hat{P}_{23},\hat{2},3)[\mathcal{A}(\hat{P}_{23},4,\hat{1})+\mathcal{A}(\hat{1},4,\hat{P}_{23})]$.
Remembering the following result for color reflection relation
\bea \mathcal{A}(1,2,...,n)=(-)^n
\mathcal{A}(n,n-1,...,1)~,~~~~\label{reverse}\eea
we know immediately that
$\mathcal{A}(\hat{P}_{23},4,\hat{1})+\mathcal{A}(\hat{1},4,\hat{P}_{23})=0$,
i.e., sum of these two terms is zero. Similarly the contribution
from the second term of (\ref{A1423}) plus the contribution from
(\ref{A1342}) is zero. We can see that adding them together we get
zero and reproduce the relation (\ref{U1-4}).

%%%%%%%%%%%%%%%%%%%%%%%
\subsection{$n$-point case}
%%%%%%%%%%%%%%%%%%%%%%%%
We want to demonstrate the $U(1)$-decoupling relation by induction.
We will prove that if $U(1)$-decoupling relation are true for all
tree level super-amplitudes less than $n$-point, then the $n$-point
relation must also be true.

The $n$-point $U(1)$-decoupling relation for $\mathcal{N}=4$ SYM can
be written as
\bea \sum_{\sigma\in~
Cyclic}\mathcal{A}(1,\sigma(2,3,\ldots,n))=0~,~~~~\eea
where leg 1 is fixed. By using cyclic relation we can always fix leg
2 instead of leg 1 and rewrite this relation in a second form
\bea \sum_{\sigma}\mathcal{A}(2,\sigma(1,3,\ldots,n))=0~,\eea
where the sum of $\sigma$ is over ordered permutations of
$\{1\}\cup\{3,\ldots,n\}$. Using BCFW recursion relation of
$\mathcal{N}=4$ SYM and $(2,n)$-shifting, we can write down all
terms of BCFW expansion. In order to show the relation among various
recursion terms clearly, here we will introduce the split sign $|$
to express one term of BCFW expansion, i.e.,
\bea \mathcal{A}(1,\ldots,i|i+1,\ldots,n)\equiv\int d^{4}\eta\
\mathcal{A}(1,\ldots,i,-P;\eta,\eta_L){1\over
P^2}\mathcal{A}(P,i+1,\ldots,n;\eta,\eta_R)~.~~~\eea
When there are more than one split sign in one amplitude, we mean
the sum of each corresponding term, for example,
\bea \mathcal{A}(1,2|3|4,5,6)&\equiv&\int d^{4}\eta \
\mathcal{A}(1,2,-P_{12};\eta,\eta_L){1\over
P_{12}^2}\mathcal{A}(P_{12},3,4,5,6;\eta,\eta_R)\nonumber\\
&&+\int d^{4}\eta \ \mathcal{A}(1,2,3,-P_{123};\eta,\eta_L){1\over
P_{123}^2}\mathcal{A}(P_{123},4,5,6;\eta,\eta_R)~.\eea
Using this notation, we can write down all terms of BCFW expansion
under $(2,n)$-shifting. Explicitly we have a special case,
\bea
\mathcal{A}(1,2,3,\ldots,n)=\mathcal{A}(1,\hat{2}|3|\ldots|n-1,\hat{n})+\mathcal{A}(\hat{2},3|4|\ldots,|\hat{n},1)~,~~~~\label{u1term1}\eea
and other general cases in which leg 2 and leg $n$ are adjacent,
such as
\begin{eqnarray}
&&\mathcal{A}(1,n,2,\ldots,n-1)=\mathcal{A}(\hat{2},3|4|\ldots|n-2|n-1|1,\hat{n})~,\\
&&\mathcal{A}(1,n-1,n,2,\ldots,n-2)=\mathcal{A}(\hat{2},3|4|\ldots|n-2|1|n-1,\hat{n})~,\end{eqnarray}
and so on. Terms of one certain super-amplitude can be divided into
two parts, characterized by the splits before leg 1 and after leg 1.

We will show that sum of all these terms equals to zero. Take the
second term in the special case (\ref{u1term1}) and all terms split
before leg 1 in general cases, and arrange them as follows,
\bean &&\mathcal{A}(\hat{2},3|1,4,5,\ldots,n-2,n-1,\hat{n}) \\
&+&\mathcal{A}(\hat{2},3|4|1,5,\ldots,n-2,n-1,\hat{n})\\
&+&\mathcal{A}(\hat{2},3|4|5|1,\ldots,n-2,n-1,\hat{n})\\
&+&\ldots\\
&+&\mathcal{A}(\hat{2},3|4|5|6|\ldots|n-2|1,n-1,\hat{n})\\
&+&\mathcal{A}(\hat{2},3|4|5|6|\ldots|n-2|n-1|1,\hat{n})\\
&+&\mathcal{A}(\hat{2},3|4|5|6|\ldots|n-2|n-1|\hat{n},1)~,\eean
then we can group all terms of the same channel $P_{2,\ldots,k}$,
i.e., terms of the same vertical line, into one part. Using cyclic
relation to fix leg 1 of all the right hand side
sub-super-amplitudes, we have
\bea \int
d^4\eta_{\hat{P}}\mathcal{A}(\hat{2},3,\ldots,k,-\hat{P}){1\over
P^2}\sum_{\sigma\in~Cyclic}
\mathcal{A}(1,\sigma(k+1,\ldots,\hat{n},\hat{P}))~.~~~\eea
It is clearly seen that sum of the same channel $P_{2,\ldots,k}$ is
zero because of $(n-k+2)$-point $U(1)$-decoupling relation. The same
argument holds for the sum of the first term in (\ref{u1term1}) and
all terms split after leg 1 in general cases. Then we have summed up
all terms and gotten a zero result, which proved the general
$U(1)$-decoupling relation.

%%%%%%%%%%%%%%%%%%%%%%%%%%%
\section{The KK relations}
%%%%%%%%%%%%%%%%%%%%%%%%%%%

Again to get some sense of these relations, let us see some special
cases. The case of $n=3$ is simplest and it is given by
\bea \mathcal{A}(1,3,\{ 2\}) = (-) \mathcal{A}(1,2,3)~.\eea
As we have mentioned before, this relation can also be considered as
color reflection relation or $U(1)$-decoupling relation of
three-point. In fact, the color reflection relation and
$U(1)$-decoupling relation are both special cases of KK relations,
connected by cyclic relation of amplitudes. We will prove the
general KK relations by induction later, and firstly let us get some
sense of these special cases.

The general KK relations of $\mathcal{N}=4$ SYM are given
by\cite{Kleiss:1988ne,BjerrumBohr:2009rd}
\bea  \mathcal{A}(1,\{\a\}, n,\{\b\}) = (-1)^{n_\b}\sum_{\sigma\in
OP(\{\a\},\{\b^T\})} \mathcal{A}(1,\sigma,
n)~.~~~~\label{KK-rel-n4}\eea
The sum is over all ordered permutations of set $\a \bigcup \b^T$,
where the relative ordering in each set  $\a$  and $\b^T$, which is
the reversed ordering of set $\b$, is preserved. The $n_\b$ is the
number of elements in set $\{\b\}$. If set $\{\beta\}$ is an empty
set, KK relations become $\mathcal{A}=\mathcal{A}$ identity.

%%%%%%%%%%%%%%%%%%%%%
\subsection{Color reflection relation as a special case}
%%%%%%%%%%%%%%%%%%%%%%

Considering the special case that $\{\alpha\}$ is an empty set, we
have
\bea
\mathcal{A}(1,n,\{\beta_1,\ldots,\beta_m\})=(-)^m\mathcal{A}(1,\beta_m,\ldots,\beta_1,n)=(-)^{m+2}\mathcal{A}(1,\beta_m,\ldots,\beta_1,n)~.~~\eea
Using cyclic relation we have
$\mathcal{A}(1,n,\beta_1,\ldots,\beta_m)=\mathcal{A}(n,\beta_1,\ldots,\beta_m,1)$,
together with above result we get the wanted $(m+2)$-point color
reflection relation.

%%%%%%%%%%%%%%%%%%%%%%
\subsection{$U(1)$-decoupling relation as a special case}
%%%%%%%%%%%%%%%%%%%%%%
$U(1)$-decoupling relation is also a special case of KK relations if
we consider that set $\{\alpha\}$ or $\{\beta\}$ has only one
element. In the case that set $\{\beta\}$ has one element, we have
\bea
\mathcal{A}(1,\{\alpha_1,\ldots,\alpha_k\},n,\beta)=-\sum_{i=0}^{k}\mathcal{A}(1,\alpha_1,\ldots,\alpha_i,\beta,\alpha_{i+1},\ldots,\alpha_k,n)~.~~\eea
Using cyclic relation we could fix leg $\beta$ instead of leg 1,
then we have
\bea \sum_{\sigma\in ~Cyclic}
\mathcal{A}(\beta,\sigma(n,1,\alpha_1,\ldots,\alpha_k))=0~.~~\eea
This is the $(k+3)$-point $U(1)$-decoupling relation.

Considering set $\{\alpha\}$ with only one element, we have
\bea
\mathcal{A}(1,\alpha,n,\{\beta_1,\ldots,\beta_m\})=(-)^{m}\sum_{i=0}^{m}
\mathcal{A}(1,\beta_m,\ldots,\beta_{i+1},\alpha,\beta_i,\ldots,\beta_1,n)~.~~\eea
Using the $(m+3)$-point color reflection relation for right hand
side, we have
\bea
\mathcal{A}(1,\alpha,n,\{\beta_1,\ldots,\beta_m\})=(-)^{2m+3}\sum_{i=0}^{m}
\mathcal{A}(n,\beta_1,\ldots,\beta_{i},\alpha,\beta_{i+1},\ldots,\beta_m,1)~.~~\eea
Using cyclic relation we can again rewrite this relation in standard
form, which is nothing but $(m+3)$-point $U(1)$-decoupling relation
with $\alpha$ fixed and the cyclic ordering of
$(1,n,\beta_1,\ldots,\beta_m)$.

%%%%%%%%%%%%%%%%%%%%%%
\subsection{The proof of general case}
%%%%%%%%%%%%%%%%%%%%%%

After considering above special cases, let us come into more complex
situation, the general KK relations with both set $\{\a\}$ and
$\{\b\}$ having more than one element. The idea of demonstration is
simply the same as we have done before, using BCFW recursion
relation of SYM version to expand super-amplitudes into sum of
sub-super-amplitudes. The three-point case can be verified directly.
Let us assume that the super-amplitude is
$\mathcal{A}(1,\{\a_1,...,\a_k\}, n,\{\b_1,...,\b_m\})$, then if we
take $(1,n)$-shifting, by BCFW recursion relation of $\mathcal{N}=4$
SYM theory we could write the left hand side of KK relations
(\ref{KK-rel-n4}) as
\bea & & \mathcal{A}(1,\{\a_1,...,\a_k\}, n,\{\b_1,...,\b_m\}) ~~~~\label{KK-gen-left}\\
& = & \int d^4\eta_{\hat{P}}\left[\sum_{i=0}^k \sum_{j=0}^m
\mathcal{A}(\b_{j+1},...,\b_m,\hat{1},\a_1,...,\a_i,
\hat{P}_{ij};\eta_{\hat{P}}) {1\over P_{ij}^2}
\mathcal{A}(-\hat{P}_{ij}, \a_{i+1},...,\a_k,
\hat{n},\b_1,...,\b_j;\eta_{\hat{P}}) \right]_{(i,j)\neq
(0,m),(k,0)}\nonumber \eea
where two cases $(i=0,j=m)$ and $(i=k,j=0)$ should be excluded from
the summation. Now we use the induction for each sub-super-amplitude
\bea \mathcal{A}(\b_{j+1},...,\b_m,\hat{1},\a_1,...,\a_i,
\hat{P}_{ij};\eta_{\hat{P}}) & = & (-)^{m-j}
\sum_{\sigma_{ij}} \mathcal{A}(\hat{1}, \sigma_{ij}, \hat{P}_{ij};\eta_{\hat{P}})~,~~~\label{KK-gen-piece-1}\\
 \mathcal{A}(-\hat{P}_{ij}, \a_{i+1},...,\a_k, \hat{n},\b_1,...,\b_j;\eta_{\hat{P}}) & = & (-)^j \sum_{\W\sigma_{ij}}
 \mathcal{A}(-\hat{P}_{ij}, \W\sigma_{ij}, \hat{n};\eta_{\hat{P}})~.~~~\label{KK-gen-piece-2}\eea
The BCFW recursion of SYM version for right hand side of KK
relations (\ref{KK-rel-n4}) is
\bea \int
d^4\eta_{P_c}\sum_{\sigma_c}\sum_{c=1}^{m+k-1}\mathcal{A}(\hat{1},\gamma_1,\ldots,\gamma_c,\hat{P}_c;\eta_{\hat{P}_c}){1\over
P_{c}^2}
\mathcal{A}(-\hat{P}_{c},\gamma_{c+1},\ldots,\gamma_{m+k},\hat{n};\eta_{\hat{P}_c})~,~~\label{KK-gen-right}\eea
where
$\sigma_c=\{\gamma_1,\ldots,\gamma_{m+k}\}=OP\{\alpha\}\cup\{\beta^T\}$.

It is easy to see that for given $\sigma_{ij},\W\sigma_{ij}$,  the
combination of set $\{1,\sigma_{ij},\W\sigma_{ij},n\}$ is one
allowed permutation $\sigma_c$ of expression (\ref{KK-gen-right}).
Also the index $i,j$ specify a particular BCFW-splitting of
$\sigma_c$. In other words, we have shown that each term in
(\ref{KK-gen-left}) will be found in (\ref{KK-gen-right}). More
specifically, for each term in
$\sum_{i=0}^k\sum_{j=0}^m\sum_{\sigma_{ij}}\sum_{\tilde{\sigma}_ij}$
there is one corresponding term in
$\sum_{c=1}^{m-k+1}\sum_{\sigma_c}$ and vice versa. Considering one
fixed split in (\ref{KK-gen-right}), i.e.,
\bea \int d^4\eta_{P_c}\sum_\sigma
\mathcal{A}(\hat{1},\gamma_1,\ldots,\gamma_c,\hat{P}_c;\eta_{\hat{P}_c}){1\over
P_c^2}\mathcal{A}(-\hat{P}_c,\gamma_{c+1},\ldots,\gamma_{m+k},\hat{n};\eta_{\hat{P}_c})~,~~\label{KK-gen-r1}\eea
the number of $\gamma$ in the left hand side is $c$. Then we should
re-group terms of BCFW expansion of (\ref{KK-gen-left}) as follows.
Take the front $i'$ elements of $\{\alpha\}$ and the front $j'$
elements of $\{\beta^T\}$ which satisfies $i'+j'=c$, their ordered
permutations give $\sigma_{ij}$ if we identify $i=i', j=m-j'$. Since
$i$ can take the value from 0 to $k$ and $j$ from 0 to $m$, there
are many combinations of $(i',j')$ which satisfy $i'+j'=c$. We
should group terms of BCFW expansion according to $c$, i.e., we
group terms in (\ref{KK-gen-left}) which satisfy $i'+j'=i+m-j=c$,
and this transfers $\sum_{i=0}^k\sum_{j=0}^m$ to
$\sum_{c=1}^{m+k-1}$ and $\sigma_{ij}$ to
$\sum_{i=0}^cOP\{\alpha_1,\ldots,\alpha_i\}\cup\{\beta_m,\dots,\beta_{i+m-c+1}\}$.
For one certain $c$, this is just set $\{\gamma_1,\dots,\gamma_c\}$
in (\ref{KK-gen-r1}), where $\gamma$ takes the value from front $i'$
elements of $\{\alpha\}$ and front $j'$ elements of $\{\beta^T\}$
which satisfies $i'+j'=c$. Then terms of BCFW expansion of left hand
side and right hand side match to each other. Thus if we can show
the total number of terms is same for both (\ref{KK-gen-left}) and
(\ref{KK-gen-right}), we have proved the identity.

To count terms it is easy to see that there are
$C^i_{i+m-j}={(i+m-j)!\over i! (m-j)!}$ terms at the right hand side
of equation (\ref{KK-gen-piece-1}), while there are
$C^j_{j+k-i}={(j+k-i)!\over j! (k-i)!}$ terms at the right hand side
of equation (\ref{KK-gen-piece-2}). Thus the total number of terms
of (\ref{KK-gen-left}) is
\bea -2 {(m+k)!\over m! k!}+\sum_{i=0}^k \sum_{j=0}^m {(i+m-j)!\over
i! (m-j)!}{(j+k-i)!\over j! (k-i)!}~,~~\eea
where $-2{(m+k)!\over m! k!}$ counts the two excluded cases. The
right hand side of KK-relations will have
\bea {(k+m)!\over k! m!} (k+m-1) \eea
terms after using the BCFW recursion relation to expand each
super-amplitude into $(k+m-1)$ terms as in (\ref{KK-gen-right}).
These two numbers match up as it should be, which can be easily
checked in Mathematica.

%%%%%%%%%%%%%%%%%%%%%%%
\section{The BCJ relations}
%%%%%%%%%%%%%%%%%%%%%%%%%

%%%%%%%%%%%%%%%%%%%
\subsection{Direct verification of four-point case}
%%%%%%%%%%%%%%%%%%%%

Let us again start with the simplest case of BCJ relations, i.e.,
the $n=4$ case. There are two independent relations:
\bea s_{23} \mathcal{A}(1,2,3,4)= s_{13}
\mathcal{A}(1,3,4,2)~,~~~~s_{12} \mathcal{A}(1,2,3,4)= s_{13}
\mathcal{A}(1,4,2,3)~,~~~~\label{BCG-n=4-1}\eea
where we use the notation $s_{ij}=(k_i+k_j)^2$. By BCFW recursion
relation of $\mathcal{N}=4$ SYM theory we know that super-amplitude
of four-point can be expressed as products of two super-amplitudes
of three-point. Due to the three-point kinematics one of
$\mbox{MHV}_3$ and $\overline{\mbox{MHV}}_3$ should vanishes, so
there are only $\mbox{MHV}$ amplitudes of four-point case. Since
there is a simple function for all $\mbox{MHV}$
amplitudes\cite{Nair:1988bq}:
\bea \mathcal{A}(1,\ldots,n;\eta_1,\ldots,\eta_n)={\delta^{4}(\sum
p)\delta^{2\mathcal{N}}(\sum_i\lambda_i\eta_i)\over
\Spaa{1~2}\Spaa{2~3}\cdots\Spaa{n~1}}~,\eea
thus we can use this to directly verify relation (\ref{BCG-n=4-1}).
Take the first relation of (\ref{BCG-n=4-1}) for example, by writing
$s_{ij}=\Spaa{i~j}\Spbb{i~j}$ we have
\bea s_{23} \mathcal{A}(1,2,3,4)- s_{13}
\mathcal{A}(1,3,4,2)={\Spab{4|2|3}+\Spab{4|1|3}\over
\Spaa{1~2}\Spaa{3~4}\Spaa{4~1}\Spaa{4~2}}\delta^4(\sum
p)\delta^8(\sum_i\lambda_i\eta_i)~,\eea
and the numerator equals to zero because of energy-momentum
conservation. With simple calculation we can verify other relations
similarly.

%%%%%%%%%%%%%%%%%%%%%%
\subsection{The fundamental relation}
%%%%%%%%%%%%%%%%%%%%%%

In $\mathcal{N}=4$ SYM theory we write the general formula of BCJ
relations as\cite{Bern:2008qj,Sondergaard:2009za}
\bea \mathcal{A}(1,2,\{ 4,5,...,m\}, 3, \{ m+1,m+2,...,n\}) & = &
\sum_{\sigma_i \in POP } \mathcal{A} (1,2,3,\sigma_i) {\cal
F}~.~~~~\label{BCJ-general}\eea
The primary one is the one with $m=4$ and others could be derived by
repeatedly using this one and KK relations. For example considering
BCJ relations of five-point
case\cite{Bern:2008qj,Sondergaard:2009za}, we have the primary one
\bea 0 & = & -s_{24} \mathcal{A}(1,2,4,3,5)+(s_{14}+s_{45})
\mathcal{A}(1,2,3,4,5)+ s_{14}
\mathcal{A}(1,2,3,5,4)~.~~~~\label{BCJ-n=5-4.27}\eea
We want to show that relation (\ref{BCJ-n=5-4.27}) is the essential
one, i.e., from this we can derive all other equations. Let us try
to derive other relations, for example,
\bea  \mathcal{A}(1,4,2,3,5) & = & { -s_{12} s_{45}
\mathcal{A}(1,2,3,4,5)+s_{25}(s_{14}+s_{24})\mathcal{A}(1,4,3,2,5)\over
s_{35} s_{24}}~,\label{BCJ-n=5-6.6}\nn
\mathcal{A}(1,2,4,3,5) & = & {s_{45}(s_{12}+s_{24})
\mathcal{A}(1,2,3,4,5)-s_{25}s_{14} \mathcal{A}(1,4,3,2,5)\over
s_{35} s_{24}}~. \eea
Starting from (\ref{BCJ-n=5-4.27}) and related KK relations
\bea s_{24} \mathcal{A}(1,2,4,3,5) & = & (s_{14}+s_{45})
\mathcal{A}(1,2,3,4,5)+ s_{14}
\mathcal{A}(1,2,3,5,4)~,\\
\mathcal{A}(1,2,3,5,4) & = &
-\mathcal{A}(1,2,3,4,5)-\mathcal{A}(1,2,4,3,5)-\mathcal{A}(1,4,2,3,5)~,\eea
we can derive
\bea (s_{24}+s_{14}) \mathcal{A}(1,2,4,3,5) & = & s_{45}
\mathcal{A}(1,2,3,4,5)- s_{14}
\mathcal{A}(1,4,2,3,5)~.~~~~\label{BCJ-n=5-main-1}\eea
The advantage of this relation is that $1,5$ have been put at the
beginning and end position. Using this one, we can reduce the basis
from $(n-2)!$ to $(n-3)!$. Exchanging $2,4$ in
(\ref{BCJ-n=5-main-1}) we obtain
\bea (s_{24}+s_{12}) \mathcal{A}(1,4,2,3,5) & = & s_{25}
\mathcal{A}(1,4,3,2,5)- s_{12}
\mathcal{A}(1,2,4,3,5)~.~~~~\label{BCJ-n=5-main-2}\eea
Combining above one with (\ref{BCJ-n=5-main-1}) we can get the
wanted relations (\ref{BCJ-n=5-6.6}) immediately.

For another example we consider the following BCJ relation
\bea s_{24} s_{13} \mathcal{A}(1,2,4,5,3)= -s_{34}s_{51}
\mathcal{A}(1,2,3,4,5)-s_{14}(s_{13}+s_{35})
\mathcal{A}(1,2,3,5,4)~.\eea
To show this, we write down (\ref{BCJ-n=5-4.27}) and the one given
by the $3\leftrightarrow 4$ changing, i.e.,
\bea s_{24} \mathcal{A}(1,2,4,3,5) & = & (s_{14}+s_{45})
\mathcal{A}(1,2,3,4,5)+ s_{14}
\mathcal{A}(1,2,3,5,4)~,\\
s_{23} \mathcal{A}(1,2,3,4,5) & = & (s_{13}+s_{35})
\mathcal{A}(1,2,4,3,5)+ s_{13} \mathcal{A}(1,2,4,5,3)~.\eea
Thus we have
\bean \mathcal{A}(1,2,4,5,3) & = & { s_{23} \mathcal{A}(1,2,3,4,5) -
(s_{13}+s_{35}) \mathcal{A}(1,2,4,3,5)\over s_{13}} \nn
& = & { s_{23} \mathcal{A}(1,2,3,4,5)\over s_{13}} -
{(s_{13}+s_{35}) \over s_{13}}{(s_{14}+s_{45})
\mathcal{A}(1,2,3,4,5)+ s_{14} \mathcal{A}(1,2,3,5,4)\over
s_{24}}\nn
& = & \mathcal{A}(1,2,3,4,5) { s_{24}s_{23}-(s_{13}+s_{35})
(s_{14}+s_{45})\over s_{13} s_{24}}- \mathcal{A}(1,2,3,5,4)
{s_{14}(s_{13}+s_{35})\over s_{13} s_{24}}\nn
& = & \mathcal{A}(1,2,3,4,5) { s_{24}s_{23}-(s_{24}-s_{51})
(s_{13}-s_{51})\over s_{13} s_{24}}- \mathcal{A}(1,2,3,5,4)
{s_{14}(s_{13}+s_{35})\over s_{13} s_{24}}\nn
& = & \mathcal{A}(1,2,3,4,5) { -s_{51} s_{34}\over s_{13} s_{24}}-
\mathcal{A}(1,2,3,5,4) {s_{14}(s_{13}+s_{35})\over s_{13}
s_{24}}~.\eean
Other BCJ relations should be derived similarly from the primary one
and KK relations. Thus in the proof we could only consider the
primary one $m=4$.

%%%%%%%%%%%%%%%%%%%%%%%%
\subsection{The primary formula of BCJ relation}
%%%%%%%%%%%%%%%%%%%%%%%

For the special case of BCJ relations when $m=4$, the general
formula (\ref{BCJ-general}) is given by
\bea s_{24}\mathcal{A}(1,2,4,3,5,6,...,n) & = &
\mathcal{A}(1,2,3,4,5,6,...,n)( s_{41}+ s_{45}+...+s_{4n})\nn & & +
\sum_{i=5}^n \mathcal{A}(1,2,3,5,...,i,4,i+1,...,n)
(s_{41}+\sum_{k=i+1}^n s_{4k})~.~\eea
In other words, we have leg $4$ in all possible positions inserted
into string $(5,6,...,n)$ and sum up $s_{41}$ and all $s_{4t}$ with
number $t$ at the right hand side of leg $4$. For expression
simplicity we re-mark the legs as
$(2,4,3,5,6,\ldots,n,1)\to(1,2,3,4,5,\ldots,n-1,n)$, then we have
\bea \sum_{i=3}^n \left[\mathcal{A}(1,3,\ldots,i-1,2,i,\ldots,n-1,n)
\sum_{j=i}^n s_{2j} \right]\equiv I_n =0~, ~~~~\label{gbcj}\eea
which means that we have leg $2$ in all possible positions inserted
into string $(3,4,\ldots,n-1)$ and sum up $s_{2j}$ with number $j$
at the right hand side of leg $2$. Note that $\sum_{j=1}^n
s_{2j}=0$, i.e., $\sum_{j=i}^n s_{2j}=-\sum_{j=1}^{i-1} s_{2j}$,
thus we have a dual form of BCJ relations
\bea \sum_{i=3}^n \left[\mathcal{A}(1,3,\ldots,i-1,2,i,\ldots,n-1,n)
\sum_{j=1}^{i-1} s_{2j} \right] &=&0~,~~ \eea
with number $j$ at the left hand side of leg $2$.

Since the lower point BCJ relations can be checked directly, we
focus on the $n$-point relations and assume that BCJ relations are
true for super-amplitudes lower than $n$-point. We use BCFW
recursion of SYM theory to expand all super-amplitudes in
(\ref{gbcj}) under $(1,n)$-shifting, such as
\bea &&\mathcal{A}(1,3,\ldots,i-1,2,i,\ldots,n-1,n)\nn
&&=\sum_{k=3}^{i-2} \int d^4 \eta_{\hat{P}_k}
\mathcal{A}(\hat{1},3,\ldots,k,\hat{P}_k;\eta_{\hat{P}_k}){1\over
P_k^2}\mathcal{A}(-\hat{P}_k,k+1,\ldots,i-1,2,i,\ldots,n-1,\hat{n};\eta_{\hat{P}_k})\nn
&&\phantom{\sum}+\int d^4 \eta_{\hat{P}_{i-1}}
\mathcal{A}(\hat{1},3,\ldots,i-1,\hat{P}_{i-1};\eta_{\hat{P}_{i-1}}){1\over
P_{i-1}^2}\mathcal{A}(-\hat{P}_{i-1},2,i,\ldots,n-1,\hat{n};\eta_{\hat{P}_{i-1}})\bigg|_{i\neq3}\nn
&&\phantom{\sum}+ \int d^4 \eta_{\hat{P}_2}
\mathcal{A}(\hat{1},3,\ldots,i-1,2,\hat{P}_2;\eta_{\hat{P}_2}){1\over
P_2^2}\mathcal{A}(-\hat{P}_2,i,\ldots,n-1,\hat{n};\eta_{\hat{P}_2})\bigg|_{i\neq
n}\nn &&+\sum_{k=i}^{n-2}\int d^4 \eta_{\hat{P}_k}
\mathcal{A}(\hat{1},3,\ldots,i-1,2,i,\ldots,k,\hat{P}_k;\eta_{\hat{P}_k}){1\over
P_k^2}\mathcal{A}(-\hat{P}_k,k+1,\ldots,n-1,\hat{n};\eta_{\hat{P}_k})\nn
&&\equiv A_i+B_i+C_i+D_i~.~~ \label{abcfw} \eea
$A_i$ is terms split at the left hand side of leg $(i-1)$ and $D_i$
is terms at the right hand side of leg $i$. $B_i$ and $C_i$ are
special terms which do not exist when $i$ takes the boundary value
$i=3$ or $n$. In later discussion we denote
$A=\sum_{i=3}^n(A_i\sum_{j=i}^ns_{2j})$, which is the corresponding
contribution of all the super-amplitudes, so for $B,C,D$. The left
hand side of (\ref{gbcj}) after BCFW expansion is $I_n=A+B+C+D$.
Firstly we consider the term $A$, which is
\bea \sum_{i=3}^n \sum_{k=3}^{i-2} \left[\int d^4 \eta_{\hat{P}_k}
\mathcal{A}(\hat{1},3,\ldots,k,\hat{P}_k;\eta_{P_k}){1\over
P_k^2}\mathcal{A}(-\hat{P}_k,k+1,\ldots,i-1,2,i,\ldots,n-1,\hat{n};\eta_{P_k})\sum_{j=i}^n
s_{2j}\right]~,\nn\eea
where propagator is distinguished by index $k$ and index $i$ denotes
distinct super-amplitude. In order to group terms of same channel
into one super-space integration $\eta_{\hat{P}_k}$, we should
exchange the order of summation from $\sum_{i=3}^n \sum_{k=3}^{i-2}$
to $\sum_{k=3}^{n-2} \sum_{i=k+2}^{n}$, i.e.,
\bea \sum_{k=3}^{n-2}& &  \int d^4 \eta_{\hat{P}_k}
\bigg\{\mathcal{A}(\hat{1},3,\ldots,k,\hat{P}_k;\eta_{\hat{P}_k}){1\over
P_k^2}\times\nn && \bigg[\sum_{i=k+2}^{n}
\bigg(\mathcal{A}(-\hat{P}_k,k+1,\ldots,i-1,2,i,\ldots,n-1,\hat{n};\eta_{\hat{P}_k})\sum_{j=i}^n
s_{2j}\bigg)\bigg]\bigg\}~,~~\label{1term}
 \eea
where propagator $P_{k}$ is related by index $k$, so is the shifted
Grassmann variable $\eta_{\hat{P}_{k}}$. All $z$ in super-amplitudes
and kinematic factors $s_{2j}$ should be replaced by $z_P$, which
are fixed by on-shell equation $\hat{P}^2(z)=0$. Now let us consider
the term in square brackets. It is noticed that in the right hand
side sub-super-amplitudes every momentum is on-shell and the shifted
momenta become $\hat{P}_k=\hat{P}_k(z_{P_k})$ and
$\hat{n}=\hat{n}(z_{P_k})$, but $\sum_{j=i}^ns_{2j}$ is evaluated
with un-shifted momenta. In order to use lower point BCJ relations
we should rewrite the factor $\sum_{j=i}^n s_{2j}$ as
$\sum_{j=i}^{\hat{n}}
s_{2j}(z_{P_k})+(s_{2n}-s_{2\hat{n}}(z_{P_k}))$. Note that all $z$
in these factors have been replaced by $z_{P_k}$, thus the term in
square brackets is split into two parts,
\bea &&\sum_{i=k+2}^{n}
\left(\mathcal{A}(-\hat{P}_k,k+1,\ldots,i-1,2,i,\ldots,n-1,\hat{n};\eta_{\hat{P}_k})\sum_{j=i}^n
s_{2j}\right)\nn &&=\sum_{i=k+2}^{n}
\left(\mathcal{A}(-\hat{P}_k,k+1,\ldots,i-1,2,i,\ldots,n-1,\hat{n};\eta_{\hat{P}_k})\sum_{j=i}^{\hat{n}}
s_{2j}\right)\nn &&\phantom{=}+(s_{2n}-s_{2\hat{n}})\sum_{i=k+2}^{n}
\mathcal{A}(-\hat{P}_k,k+1,\ldots,i-1,2,i,\ldots,n-1,\hat{n};\eta_{\hat{P}_k})~.~\eea
Using $(n-k+1)$-point BCJ relation the first term becomes
\bea
-\mathcal{A}(-\hat{P}_k,2,k+1,\ldots,n-1,\hat{n};\eta_{\hat{P}_k})\sum_{j=k+1}^{\hat{n}}
s_{2j}~,\nonumber\eea
and using $(n-k+1)$-point $U(1)$-decoupling relation the second term
turns to
\bea
(s_{2\hat{n}}-s_{2n})\mathcal{A}(-\hat{P}_k,k+1,\ldots,n-1,\hat{n},2;\eta_{\hat{P}_k})~.~~\nonumber\eea
After pulling them back into term $A$ and changing index $k$ to
$i-1$, we combine them with term B. Many terms naturally cancel and
we have the result of $A+B$,
\bea \sum_{i=4}^{n}\left[\int d^4 \eta_{\hat{P}_{i-1}}
\mathcal{A}(\hat{1},3,\ldots,i-1,\hat{P}_{i-1};\eta_{\hat{P}_{i-1}}){1\over
P_{i-1}^2}
\mathcal{A}(-\hat{P}_{i-1},i,\ldots,n-1,\hat{n},2;\eta_{\hat{P}_{i-1}})(s_{2\hat{n}}(z_{P_{i-1}})-s_{2n})\right]~.~\nonumber\\\label{ab}\eea

The other two terms in (\ref{abcfw}), $C$ and $D$, are disposed in
almost the same manner, while we change the kinematic factors
$\sum_{j=i}^{\hat{n}} s_{2j}$ to $-\sum_{j=\hat{1}}^{i-1} s_{2j}$ in
order to use the dual form of BCJ relations. In order to use BCJ
relations of lower point, we should rewrite $\sum_{j=1}^{i-1}
s_{2j}$ as $\sum_{j=\hat{1}}^{i-1} s_{2j}+(s_{21}-s_{2\hat{1}})$.
Notice the important property $\hat{P}_1(z)+\hat{P}_n(z)=P_1+P_n$ of
BCFW recursion relation, we have
$s_{2\hat{n}}-s_{2n}=-(s_{2\hat{1}}-s_{21})$. With this in hand we
can easily repeat calculation of $C+D$, which gives the result
\bea \sum_{i=3}^{n-1}\left[\int d^4 \eta_{\hat{P}_{i-1}}
\mathcal{A}(2,\hat{1},3,\ldots,i-1,\hat{P}_{i-1};\eta_{\hat{P}_{i-1}}){1\over
P_{i-1}^2}
\mathcal{A}(-\hat{P}_{i-1},i,\ldots,n-1,\hat{n};\eta_{\hat{P}_{i-1}})(s_{2\hat{n}}(z_{P_{i-1}})-s_{2n})\right]~.~\label{cd}\nn
\eea

Experience in doing BCFW recursion relation indicates that
$I_n=A+B+C+D$ is related to the supersymmetric BCFW recursion
relation of $\mathcal{A}(2,1,3,\ldots,n)$ under $(1,n)$-shifting. In
fact let us consider integration
\bea \oint {dz\over z}
\mathcal{A}(2,\hat{1},3,\ldots,n-1,\hat{n})\times
s_{2\hat{n}}(z)~.~~~~\label{bcjInt}\eea
Residue at pole $z=0$ gives
\bean \mathcal{A}(2,1,3,\ldots,n)s_{2n}~,~~\eean
and sum of poles at shifted super-amplitude gives
\bean -\sum_{i=4}^{n}\left[\int d^4 \eta_{\hat{P}_{i-1}}
\mathcal{A}(\hat{1},3,\ldots,i-1,\hat{P}_{i-1};\eta_{\hat{P}_{i-1}}){1\over
P_{i-1}^2}
\mathcal{A}(-\hat{P}_{i-1},i,\ldots,n-1,\hat{n},2;\eta_{\hat{P}_{i-1}})s_{2\hat{n}}(z_{P_{i-1}})\right]~,~\eean
and
\bean -\sum_{i=3}^{n-1}\left[\int d^4 \eta_{\hat{P}_{i-1}}
\mathcal{A}(2,\hat{1},3,\ldots,i-1,\hat{P}_{i-1};\eta_{\hat{P}_{i-1}}){1\over
P_{i-1}^2}
\mathcal{A}(-\hat{P}_{i-1},i,\ldots,n-1,\hat{n};\eta_{\hat{P}_{i-1}})s_{2\hat{n}}(z_{P_{i-1}})\right]~.~
\eean
Adding these terms together we reproduce $-I_n$. Since the shifted
momenta $(1,n)$ are not adjacent, the super-amplitude behaves as
$1/z^2$ when $z$ approaches to infinity\cite{ArkaniHamed:2008gz}, so
that the integration (\ref{bcjInt}) equals to zero. Thus we have
$I_n=0$, which finishes the proof.

\section{Conclusion}

It is interesting to see that in $\mathcal{N}=4$ SYM theory there
are same relations of amplitudes as pure gluon
amplitudes\cite{Kleiss:1988ne,Bern:2008qj,BjerrumBohr:2009rd}, and
one can prove them using BCFW recursion relation of $\mathcal{N}=4$
SYM theory, like the pure gluon case\cite{Feng:2010my}. This
verifies the conjecture that matter amplitudes also obey the similar
BCJ relations\cite{Sondergaard:2009za}. Deduced from relations
between amplitudes of $\mathcal{N}=4$ SYM theory and amplitudes of
$\mathcal{N}=8$ super-gravity, it is challenging and interesting to
think whether one can use supersymmetric BCFW recursion relation to
guess or prove relations of super-gravity amplitudes. Since BCFW
recursion relation naturally groups many terms into different
channels, we wander if it could shed some light on computing gravity
amplitudes through square relations of BCJ.

%%%%%%%%%%%%%%%%%%%%%%%%%%%%%%%%%
\section*{Acknowledgments}
%%%%%%%%%%%%%%%%%%%%%%%%%%%%%%%%%
We would like to thank Prof. Bo Feng for helpful discussions. This
work is supported by fund from Qiu-Shi, the Fundamental Research
Funds for the Central Universities with contract number 2009QNA3015,
as well as Chinese NSF funding under contract No.10875104.

%%%%%%%%%%%%%%%%%%%%%%%%%%%%%%%%%%%%%%%%%%%%%%%%%%%%


\begin{thebibliography}{999}



%\cite{Britto:2004ap}
\bibitem{Britto:2004ap}
  R.~Britto, F.~Cachazo and B.~Feng,
  %``New Recursion Relations for Tree Amplitudes of Gluons,''
  Nucl.\ Phys.\  B {\bf 715}, 499 (2005)
  [arXiv:hep-th/0412308].
  %%CITATION = NUPHA,B715,499;%%

%\cite{Britto:2005fq}
\bibitem{Britto:2005fq}
  R.~Britto, F.~Cachazo, B.~Feng and E.~Witten,
  %``Direct Proof Of Tree-Level Recursion Relation In Yang-Mills Theory,''
  Phys.\ Rev.\ Lett.\  {\bf 94}, 181602 (2005)
  [arXiv:hep-th/0501052].
  %%CITATION = PRLTA,94,181602;%%

%\cite{Witten:2003nn}
\bibitem{Witten:2003nn}
  E.~Witten,
  %``Perturbative gauge theory as a string theory in twistor space,''
  Commun.\ Math.\ Phys.\  {\bf 252}, 189 (2004)
  [arXiv:hep-th/0312171].
  %%CITATION = CMPHA,252,189;%%


%\cite{Bianchi:2008pu}
\bibitem{Bianchi:2008pu}
  M.~Bianchi, H.~Elvang and D.~Z.~Freedman,
  %``Generating Tree Amplitudes in N=4 SYM and N = 8 SG,''
  JHEP {\bf 0809}, 063 (2008)
  [arXiv:0805.0757 [hep-th]].
  %%CITATION = JHEPA,0809,063;%%


%\cite{Brandhuber:2008pf}
\bibitem{Brandhuber:2008pf}
  A.~Brandhuber, P.~Heslop and G.~Travaglini,
  %``A note on dual superconformal symmetry of the N=4 super Yang-Mills
  %S-matrix,''
  Phys.\ Rev.\  D {\bf 78}, 125005 (2008)
  [arXiv:0807.4097 [hep-th]].
  %%CITATION = PHRVA,D78,125005;%%

%\cite{ArkaniHamed:2008gz}
\bibitem{ArkaniHamed:2008gz}
  N.~Arkani-Hamed, F.~Cachazo and J.~Kaplan,
  %``What is the Simplest Quantum Field Theory?,''
  arXiv:0808.1446 [hep-th].
  %%CITATION = ARXIV:0808.1446;%%

%\cite{Elvang:2008na}
\bibitem{Elvang:2008na}
  H.~Elvang, D.~Z.~Freedman and M.~Kiermaier,
  %``Recursion Relations, Generating Functions, and Unitarity Sums in N=4 SYM
  %Theory,''
  JHEP {\bf 0904}, 009 (2009)
  [arXiv:0808.1720 [hep-th]].
  %%CITATION = JHEPA,0904,009;%%



%\cite{Drummond:2008cr}
\bibitem{Drummond:2008cr}
  J.~M.~Drummond and J.~M.~Henn,
  %``All tree-level amplitudes in N=4 SYM,''
  JHEP {\bf 0904}, 018 (2009)
  [arXiv:0808.2475 [hep-th]].
  %%CITATION = JHEPA,0904,018;%%


%\cite{Kleiss:1988ne}
\bibitem{Kleiss:1988ne}
  R.~Kleiss and H.~Kuijf,
  %``MULTI - GLUON CROSS-SECTIONS AND FIVE JET PRODUCTION AT HADRON COLLIDERS,''
  Nucl.\ Phys.\  B {\bf 312}, 616 (1989).
  %%CITATION = NUPHA,B312,616;%%

%\cite{Bern:2008qj}
\bibitem{Bern:2008qj}
  Z.~Bern, J.~J.~M.~Carrasco and H.~Johansson,
  %``New Relations for Gauge-Theory Amplitudes,''
  Phys.\ Rev.\  D {\bf 78}, 085011 (2008)
  [arXiv:0805.3993 [hep-ph]].
  %%CITATION = PHRVA,D78,085011;%%




%\cite{DelDuca:1999rs}
\bibitem{DelDuca:1999rs}
  V.~Del Duca, L.~J.~Dixon and F.~Maltoni,
  %``New color decompositions for gauge amplitudes at tree and loop level,''
  Nucl.\ Phys.\  B {\bf 571}, 51 (2000)
  [arXiv:hep-ph/9910563].
  %%CITATION = NUPHA,B571,51;%%



%\cite{BjerrumBohr:2009rd}
\bibitem{BjerrumBohr:2009rd}
  N.~E.~J.~Bjerrum-Bohr, P.~H.~Damgaard and P.~Vanhove,
  %``Minimal Basis for Gauge Theory Amplitudes,''
  Phys.\ Rev.\ Lett.\  {\bf 103}, 161602 (2009)
  [arXiv:0907.1425 [hep-th]].
  %%CITATION = PRLTA,103,161602;%%

%\cite{Stieberger:2009hq}
\bibitem{Stieberger:2009hq}
  S.~Stieberger,
  %``Open & Closed vs. Pure Open String Disk Amplitudes,''
  arXiv:0907.2211 [hep-th].
  %%CITATION = ARXIV:0907.2211;%%


  %\cite{Tye:2010dd}
\bibitem{Tye:2010dd}
  H.~Tye and Y.~Zhang,
  %``Dual Identities inside the Gluon and the Graviton Scattering Amplitudes,''
  arXiv:1003.1732 [hep-th].
  %%CITATION = ARXIV:1003.1732;%%


%\cite{BjerrumBohr:2010zs}
\bibitem{BjerrumBohr:2010zs}
  N.~E.~J.~Bjerrum-Bohr, P.~H.~Damgaard, T.~Sondergaard and P.~Vanhove,
  %``Monodromy and Jacobi-like Relations for Color-Ordered Amplitudes,''
  arXiv:1003.2403 [hep-th].
  %%CITATION = ARXIV:1003.2403;%%

%\cite{Mafra:2009bz}
\bibitem{Mafra:2009bz}
  C.~R.~Mafra,
  %``Simplifying the Tree-level Superstring Massless Five-point Amplitude,''
  JHEP {\bf 1001}, 007 (2010)
  [arXiv:0909.5206 [hep-th]].
  %%CITATION = JHEPA,1001,007;%%



%\cite{Feng:2010my}
\bibitem{Feng:2010my}
  B.~Feng, R.~Huang and Y.~Jia,
  %``U(1)-decoupling, KK-relation and BCJ relation by BCFW relation in S-matrix
  %Program,''
  arXiv:1004.3417 [hep-th].
  %%CITATION = ARXIV:1004.3417;%%

%\cite{Benincasa:2007xk}
\bibitem{Benincasa:2007xk}
  P.~Benincasa and F.~Cachazo,
  %``Consistency Conditions on the S-Matrix of Massless Particles,''
  arXiv:0705.4305 [hep-th].
  %%CITATION = ARXIV:0705.4305;%%

%\cite{S-matrix}
\bibitem{S-matrix} D.I. Olive, Phys. Rev. 135,B 745(1964); G.F.
Chew, "The Analytic S-Matrix: A Basis for Nuclear Democracy",
W.A.Benjamin, Inc., 1966; R.J. Eden, P.V. Landshoff, D.I. Olive,
J.C. Polkinghorne, "The Analytic S-Matrix", Cambridge University
Press, 1966.

%\cite{Sondergaard:2009za}
\bibitem{Sondergaard:2009za}
  T.~Sondergaard,
  %``New Relations for Gauge-Theory Amplitudes with Matter,''
  Nucl.\ Phys.\  B {\bf 821}, 417 (2009)
  [arXiv:0903.5453 [hep-th]].
  %%CITATION = NUPHA,B821,417;%%



\bibitem{Drummond}
 J.M.Drummond, J.Henn, G.P.Korchemsky and E.Sokatchev,
% ``Dual superconformal symmetry of scattering amplitudes in
%  $\mathcal {N}$=4 super-Yang Mills theory,''
  Nucl.\ Phys.\  B {\bf 828}(2010)317
  [arXiv:0807.1095 [hep-th]].

%\cite{Nair:1988bq}
\bibitem{Nair:1988bq}
  V.~P.~Nair,
  %``A CURRENT ALGEBRA FOR SOME GAUGE THEORY AMPLITUDES,''
  Phys.\ Lett.\  B {\bf 214}, 215 (1988).
  %%CITATION = PHLTA,B214,215;%%

%\cite{ArkaniHamed:2008yf}
\bibitem{ArkaniHamed:2008yf}
  N.~Arkani-Hamed and J.~Kaplan,
  %``On Tree Amplitudes in Gauge Theory and Gravity,''
  JHEP {\bf 0804}, 076 (2008)
  [arXiv:0801.2385 [hep-th]].
  %%CITATION = JHEPA,0804,076;%%




%\cite{Vaman:2005dt}
\bibitem{Vaman:2005dt}
  D.~Vaman and Y.~P.~Yao,
  %``QCD recursion relations from the largest time equation,''
  JHEP {\bf 0604}, 030 (2006)
  [arXiv:hep-th/0512031].
  %%CITATION = JHEPA,0604,030;%%

%\cite{Draggiotis:2005wq}
\bibitem{Draggiotis:2005wq}
  P.~D.~Draggiotis, R.~H.~P.~Kleiss, A.~Lazopoulos and C.~G.~Papadopoulos,
  %``Diagrammatic proof of the BCFW recursion relation for gluon amplitudes in
  %QCD,''
  Eur.\ Phys.\ J.\  C {\bf 46}, 741 (2006)
  [arXiv:hep-ph/0511288].
  %%CITATION = EPHJA,C46,741;%%

%\cite{Benincasa:2007qj}
\bibitem{Benincasa:2007qj}
  P.~Benincasa, C.~Boucher-Veronneau and F.~Cachazo,
  %``Taming tree amplitudes in general relativity,''
  JHEP {\bf 0711}, 057 (2007)
  [arXiv:hep-th/0702032].
  %%CITATION = JHEPA,0711,057;%%



%\cite{Cheung:2008dn}
\bibitem{Cheung:2008dn}
  C.~Cheung,
  %``On-Shell Recursion Relations for Generic Theories,''
  arXiv:0808.0504 [hep-th].
  %%CITATION = ARXIV:0808.0504;%%


%\cite{Feng:2009ei}
\bibitem{Feng:2009ei}
  B.~Feng, J.~Wang, Y.~Wang and Z.~Zhang,
  %``BCFW Recursion Relation with Nonzero Boundary Contribution,''
  JHEP {\bf 1001}, 019 (2010)
  [arXiv:0911.0301 [hep-th]].
  %%CITATION = JHEPA,1001,019;%%



%\cite{Feng:2010ku}
\bibitem{Feng:2010ku}
  B.~Feng and C.~Y.~Liu,
  %``A note on the boundary contribution with bad deformation in gauge theory,''
  arXiv:1004.1282 [hep-th].
  %%CITATION = ARXIV:1004.1282;%%

%\cite{Spradlin:2008bu}
\bibitem{Spradlin:2008bu}
  M.~Spradlin, A.~Volovich and C.~Wen,
  %``Three Applications of a Bonus Relation for Gravity Amplitudes,''
  Phys.\ Lett.\  B {\bf 674}, 69 (2009)
  [arXiv:0812.4767 [hep-th]].
  %%CITATION = PHLTA,B674,69;%%


\end{thebibliography}
\end{document}